\documentclass{ceurart}

\usepackage{multirow}
\usepackage{dsfont}
\usepackage{pifont}
\usepackage{xcolor}
\usepackage{subcaption}
\usepackage{wrapfig}

\usepackage{todonotes}

\begin{document}

%%
%% Rights management information (CC-BY is the CEUR-WS default license)
\copyrightyear{2026}
\copyrightclause{Copyright for this paper by its authors.
  Use permitted under Creative Commons License Attribution 4.0
  International (CC BY 4.0).}

%%
%% Conference information — UPDATE with the actual workshop details
\conference{IntRS'26: Joint Workshop on Interfaces and Human Decision Making for Recommender Systems, September 28, 2026, Minneapolis.}

%%
%% Title
\title{Why didn't more people see it? Recommendation transparency for providers}

%%
%% Authors and affiliations
\author[1]{Meysam Varasteh}[%
email=meysam.varasteh@colorado.edu,
]
\cormark[1]
\address[1]{Department of Computer Science, University of Colorado Boulder,
  Boulder, CO 80309, USA}

\author[2]{Robin Burke}[%%orcid=0000-0001-5766-6434,
email=robin.burke@colorado.edu,
]
\address[2]{Department of Information Science, University of Colorado Boulder,
  Boulder, CO 80309, USA}
%\author[1]{Anonymous Author(s)}
%\address[1]{Anonymous Institution}
%% Corresponding author note
\cortext[1]{Corresponding author.}

%%
%% Abstract
\begin{abstract}
Transparency in recommender systems has been widely studied from the perspective of those receiving recommendations, yet the needs of item providers, the creators whose content is distributed through these platforms, remain largely unexplored. Providers often lack insight into how their items do or do not receive exposure in users' recommendation lists.
In this work, we address this gap by proposing a surrogate modeling approach to explain item exposure at a system level. Rather than explaining individual user–item pairs, we train a proxy model to approximate the exposure distribution produced by a recommender. By quantifying the contribution of each feature, we seek to explain the factors driving the recommendation model's decisions across the entire user base. We evaluate our approach on two datasets and three recommendation models. Results show that the surrogate model captures the global behavior of all three recommenders with high fidelity and that the most influential factors vary meaningfully across models and domains.
\end{abstract}

%%
%% Keywords

%%
%% Build the front matter
\maketitle

\section{Introduction}

%\todo[inline]{This section needs to talk about the different purposes / goals of explanation.}

Transparency and interpretability in recommender systems form a significant research area in both academia and industry. This area has focused heavily on the explanation of individual recommender system outputs. Such explanations in recommender systems serve multiple goals. They promote transparency by revealing how the system works, and scrutability by allowing users to correct it when wrong \cite{tintarev2022beyond}. They build trust and support effectiveness by increasing user confidence and helping them make better decisions. Additionally, explanations enhance persuasiveness by encouraging users to act on recommendations, improve efficiency by speeding up decision-making, and boost satisfaction by making the overall experience more enjoyable and easy to use \cite{tintarev2015explaining, tintarev2022beyond}.

We can describe this area of emphasis as one of \textit{local explanation}, explaining the output of an individual user–item pair: why item $i$ is recommended to user $u$. There is less emphasis on describing aggregated, system-wide behavior of the model, a \textit{global explanation}. While local explanations aim to illuminate the model's behavior for a specific user–item interaction, global analysis requires a broader view of the model's overall recommendation patterns \cite{chowdhury2022equi}.

Many of the forms of local explanation that have been proposed, especially those focused on persuasiveness, tend to drift away from explanations that are grounded in the details of a recommender's decision-making process, which are complex and difficult to communicate, and towards content-oriented features, whether or not such features play a part in recommendation. This may be appropriate for supporting item-level decision-making; however, global explanations of the type that we envision require \textit{veracity}, fidelity to the recommendation model, a property emphasized in work on counterfactual explanation \cite{varasteh2024comparative, barkan2026fidelity, verma2023recrec, ghazimatin2020prince, baklanov2025refining}.

While there are a variety of global explanations that might be considered, for our purposes, we are focusing on a generalization of the standard "why item $i$ for user $u$?" explanation ($i \rightarrow u$) to summarize recommendation behavior for an item across all users ($i \rightarrow U$). Some evidence from human-centered studies suggests that explanations of this type will be helpful in supporting providers in their interactions with recommenders \cite{devito2017algorithms, dinnissen2023amplifying, michalewicz2025composing, dinnissen2023looking}.

To ensure veracity, global explanations will need to be grounded in the structural characteristics of the recommendation data and an item's positioning within the recommendation model. In the purely collaborative models, which we study here, we know that an item's recommendation pattern is frequently influenced by such factors as item popularity \cite{abdollahpouri2019managing, abdollahpouri2020multistakeholder}, user profile density \cite{sahebi2013cross}, influential users \cite{eskandanian2019power, rashid2005influence}, and neighborhood interactions \cite{eskandanian2019power}. These structural properties form the foundation of the internal logic of CF models and represent an important yet largely overlooked source of explanations.

Building on these observations, we train a model as a surrogate to predict properties of the $i \rightarrow U$ relation: in particular, to predict the exposure of a target item in recommendation lists across the user base. These models are necessarily specific to the application domain / dataset and the recommendation model. From such models, we can derive correlations between structural features of datasets and recommendation models relative to item characteristics and offer explanations for the variations in system-level exposure across different items.

Our contributions in this work can be summarized as follows:
\begin{itemize}
    \item We show that item exposure is globally predictable from a small set of structural features, achieving $R^2=0.83$--$0.93$ across three recommendation models and two datasets.
    \item We demonstrate that dominant exposure drivers differ meaningfully across models and datasets, providing model-specific diagnostic transparency to providers.
    \item We show that in each algorithm a single feature (but not always the same one) explains $60-70\%$ of exposure variance, revealing that RS exposure is driven by a small number of critical signals.
\end{itemize}

\section{Problem Formulation}
Given a trained recommender, our goal is to explain the variance of \textit{exposure} across items: why certain items receive higher exposure than others across the entire user base in their top-k recommendations. Let $\mathcal{U} = \{u_m\}_{m=1}^M$ and $\mathcal{I} = \{i_n\}_{n=1}^N$ represent the sets of users and items, respectively. The historical interactions are captured in the binary matrix $R \in \{0, 1\}^{M \times N}$, where $r_{ui} = 1$ if an interaction exists and $0$ otherwise. Additionally, $\hat{r}_{u, i}=f(u, i |\theta)$ is the prediction value where $\theta$ represents the learned latent parameters of the model.

For each user $u$, the model generates a ranked list of size $k$ by sorting items in descending order of their predicted scores. The top-k recommendation set for user $u$ is defined as:
\begin{equation}
    L_u^{k} = \{ i \in \mathcal{I}  \mid \text{rank}(i; \hat{r}_{u}) \leq k \}
\end{equation}
where $k=10$ in all experiments.

We also define $\mathcal{U}_i^*$ as all users who received item $i$ in their top-k recommendation list:

\begin{equation}
\mathcal{U}_i^* = \{u \in \mathcal{U} ~|~ i \in L_u^{k} \}
\end{equation}

Based on this, we can define the exposure of item $ i$ as the proportion of users whose top-$k$ recommendation lists contain the item across all users' recommendation lists: $E_i = |\mathcal{U}_i^*|/|U|$.

Let $X_i$ be the vector of features (described in Section~\ref{subsec:features}) derived from the dataset for item $i$. We aim to train a surrogate model $g$ such that: $E_i \approx g(X_i)$.

The quality of this approximation can be evaluated using the Coefficient of Determination ($R^2$), which calculates the proportion of the variance accounted for by the model.

\section{Approach}
\subsection{Datasets and Experimental Setup}
%\todo[inline]{Why these recommenders?}
\begin{table}[h]
\centering
\caption{Dataset Statistics}
\label{tab:datasets}
\begin{tabular}{lrrr}
\toprule
\textbf{Dataset} & \textbf{Users} & \textbf{Items} & \textbf{Sparsity} \\
\midrule
Last.fm  & 24,631 & 6,584 & 99\% \\
ML-1M    & 6,040  & 3,953 & 96\% \\
\bottomrule
\end{tabular}
\end{table}
In this paper, we evaluate three CF models with fundamentally different learning objectives: NCF~\cite{he2017neural, varasteh2023improved} (nonlinear embedding-based), VAE~\cite{liang2018variational} (generative probabilistic), and BPR~\cite{rendle2012bpr} (matrix factorization with a pairwise ranking objective)\footnote{We use the implementations provided in RecBole \url{https://recbole.io}}, evaluated on two diverse datasets: MovieLens-1M \cite{harper2015movielens} and Last.fm \cite{bertin2011million}. These differences in learning paradigms naturally produce divergent exposure distributions across models and the datasets differ in content type (movies vs. music), enabling us to assess generalization across domains.

The interaction data is split 80/20 to train and evaluate the recommender. From the trained recommendation model, recommendation lists are generated for all users. The resulting item exposures across all users after training the model are split separately 80/20 at the item level to train and evaluate the surrogate model.

Dataset statistics are shown in Table~\ref {tab:datasets}. The code for all stages of the experiments, including data generation, model training, and evaluation, is available in the associated GitHub repository \footnote{\url{https://github.com/that-recsys-lab/ProviderSideTransparency}}.
%\subsection{Surrogate Model}

As our regression model, we employed the XGBoost Regressor \cite{chen2016xgboost}, an ensemble learning algorithm based on gradient-boosted decision trees. XGBoost was selected for its ability to capture non-linear interactions between diverse feature classes. Furthermore, its built-in feature importance mechanisms let us quantify the contribution of each category to the overall recommendation behavior. Feature importance scores correspond to the gain-based importance metric provided by XGBoost, which measures the average reduction in the loss attributable to each feature across all tree splits. We used a maximum depth of $7$, number of estimators ranging from 200 to 500 (varying per model and dataset), and a learning rate of $0.01$. Finally, in this study, we use “provider” as the stakeholder associated with an item; for example, a creator or rights holder whose content is distributed by the recommender.

\subsection{Features}
\label{subsec:features}

We incorporate features derived from established literature \cite{sahebi2013cross, eskandanian2019power, rashid2005influence, abdollahpouri2019managing}, designed to capture systemic factors that influence recommendation outcomes. The features are in three categories: frequency-based, neighborhood-based, and item-centric.

\subsubsection{Frequency-based features}
There are two features in this category: item popularity (\textbf{POP}) and item-specific profile size (\textbf{SZ}). For popularity, we use a standard definition, counting the proportion of users in the dataset who have interacted with item $i$: $POP(i)=\frac{|U_i|}{  \mathcal |U|} $. For the SZ feature, we are interested in the profile size of users who received the item $i$ in their top-K recommendations, averaging across all such users. If the item has no exposure, SZ is set to zero:
\begin{equation}
    SZ(i) = \begin{cases}
    \frac{1}{|\mathcal{U}_i^*|} \sum_{u \in \mathcal{U}_i^{*}} |P_u| & \text{if } E_i > 0 \\
    0 & \text{if } E_i = 0
    \end{cases}
\end{equation}

\subsubsection{Neighborhood-based features}
For each user $u$, we identify a neighborhood of similar users, denoted as $\mathcal{N}_u$, based on the Jaccard similarity (JS) of their profiles:
\begin{equation}
    \mathcal{N}_u = \underset{v \in \mathcal{U} \setminus \{u\}}{\text{arg top-}k} \;( \text{JS}(P_u, P_v))
\end{equation}
where $P_u$ and $P_v$ denote the profiles of users $u$ and $v$, respectively. A profile is defined as the set of items the user has previously interacted with. In the experiments described below, we set $k=10$.\footnote{We experimented with varying neighborhood sizes (k=10,20,30,40) and found that the results are not sensitive to this choice.}

This calculation enables us to define Neighborhood Density (\textbf{ND}) relative to an item: the average frequency with which the target item appears in the histories of a user's nearest neighbors.
\begin{equation}
    ND(i) =
    \begin{cases}
    \frac{1}{|\mathcal{U}_i^{*}|} \sum_{u \in \mathcal{U}_i^{*}} \frac{1}{|\mathcal{N}_u|} \sum_{v \in \mathcal{N}_u} \mathbb{I}(i \in P_v) &
    \text{if } E_i > 0 \\
0 & \text{if } E_i = 0
    \end{cases}
\end{equation}

We define a different kind of neighborhood by looking at similarities between items in terms of user overlap. If two items $i$ and $j$ are interacted with the same group of people, they are considered similar:
\begin{equation}
    \label{equ:similar_items}
    \mathcal{S}_i = \underset{j \in \mathcal{I} \setminus \{i\}}{\text{arg top-}k} \;( \text{JS}(U_i, U_j))
\end{equation}
where $U_i$ and $U_j$ are the set of users who interacted with items $i$ and $j$ respectively. As with ND, we set $k=10$.\footnote{The results in our experiments were found not to be sensitive to choice of $k$. }

With this similarity metric, we are also able to define the Related Items feature ($\textbf{RI}$), which measures the fraction of items related to the target item within similar users' profiles.
\begin{equation}
    RI(i) =
    \begin{cases}

    \frac{1}{|\mathcal{U}_i^{*}|} \sum_{u \in \mathcal{U}_i^{*}} \frac{1}{|P_u|} \sum_{e \in P_u} \mathbb{I}(e \in \mathcal{S}_i)
    &
    \text{if } E_i > 0 \\
0 & \text{if } E_i = 0
    \end{cases}
\end{equation}

\subsubsection{Item-specific features}

In both datasets, items are associated with genres. If a user shows a propensity to a certain genre or category, we might expect that items of that genre would be more likely to be recommended to them. We therefore define Genre Density (\textbf{GD}) as the average number of items within the target item's genre present in the profiles of users in $\mathcal{U}_i^*$.
    \begin{equation}
    GD(i) =
     \begin{cases}
     \frac{1}{|\mathcal{U}_i^{*}|} \sum_{u \in \mathcal{U}_i^{*}} \sum_{j \in P_u} \mathbb{I}(G(j) = G(i)) &
    \text{if } E_i > 0 \\
0 & \text{if } E_i = 0
    \end{cases}
    \end{equation}
where $G(i)$ is the genre (or set of genres) associated with item $i$ and $\mathbb{I}(\cdot)$ is an indicator function.

We have three other features, drawn from the item metadata:

\begin{itemize}
    \item \textbf{YR}: The release year of the target item, utilized as a temporal indicator of the item's age.

    \item \textbf{AE} (Last.fm only): Audio energy is a high-level acoustic feature representing the perceived intensity and activity of the track, ranging from 0 to 1.

    \item \textbf{INS} (Last.fm only): Instrumentalness represents the likelihood that a track contains no vocal content, ranging from 0 to 1.
\end{itemize}

\begin{table}[t]
\centering
\caption{Feature importance, $R^2$ for surrogate model. Dashed lines indicate unavailable features for the given dataset.}
\label{tab:surrogate_results}
\begin{tabular}{lc|cccccccc|c}
\hline
\textbf{Dataset} & \textbf{Rec} & $POP$ & $SZ$ & $YR$ & $GD$ & $ND$ & $RI$ & $AE$ & $INS$ & $R^2$  \\ \hline
        & NCF & 0.1 & 0.1 & 0 & 0 & 0.8 & 0 & - & - & 0.85  \\ \cline{2-11}
\multirow{1}{*}{ML-1M}  & VAE & 0.2 & 0.1 & 0 & 0 & 0.7 & 0 & - & - & 0.93  \\ \cline{2-11}
        & BPR & 0.5 & 0.2 & 0 & 0.1 & 0.2 & 0 & - & - & 0.83  \\ \hline
        & NCF & 0 & 0.2 & 0 & 0 & 0.4 & 0.4 & 0 & 0 & 0.87  \\ \cline{2-11}
\multirow{1}{*}{Last.fm} & VAE & 0.5 & 0.1 & 0 & 0 & 0.3 & 0.1 & 0 & 0 & 0.87  \\ \cline{2-11}
        & BPR & 0.7 & 0.1 & 0 & 0 & 0.1 & 0.1 & 0 & 0 & 0.88  \\ \hline
\end{tabular}
\end{table}
\section{Results}

Table~\ref{tab:surrogate_results} reports the regression model's performance across all model-dataset combinations. The surrogate achieves high fidelity in all settings, with $R^2$ scores ranging from 0.83 to 0.93, demonstrating that structural features effectively capture the global exposure behavior of all three recommenders across both domains. Despite this overall consistency, the contribution of individual features varies substantially across models and domains, reflecting their different learning objectives and optimization strategies.

\textbf{ML-1M dataset:} On the ML-1M dataset, Neighborhood Density (ND) is the dominant feature for both NCF and VAE ($ND=0.7-0.8$), indicating that items appearing frequently in the histories of a user's nearest neighbors are strongly favored by both models. This suggests that both NCF and VAE heavily rely on neighborhood-based features when distributing exposure on the movie domain. In contrast, BPR exhibits a notably different pattern. The popularity of the item ($POP=0.5$) becomes the dominant factor, followed by the density of neighborhoods and profile size features ($ND=0.2, SZ=0.2$). This reflects BPR's pairwise ranking objective, which is more sensitive to globally popular items than to localized neighborhood structure.\footnote{Popularity bias is a well-known characteristic of BPR; see \cite{jannach2015recommenders}.}

\textbf{Last.fm Results:} Compared to ML-1M, the Related Items feature, which was negligible on ML-1M, appears to play a somewhat more notable role on Last.fm, highlighting the impact of item similarity within user profiles on the global behavior of the model in the music domain.
For NCF, the contribution is distributed across multiple features: Related Items ($RI=0.4$) and Neighborhood Density ($ND=0.4$) are the dominant factors, with average profile size contributing to a lesser extent and popularity contributing negligibly. This indicates that in the music domain, NCF relies more heavily on item similarity and collaborative neighborhood structure than on global popularity signals, favoring songs that are similar to those already present in the user's listening history and that appear frequently in the profiles of similar users.
VAE, on the other hand, exhibits a strong popularity bias ($POP=0.5$), making item popularity the dominant driver of exposure with a smaller but non-negligible contribution from neighborhood density ($ND=0.3$).
Finally, BPR exhibits a similar pattern to its behavior on ML-1M. Popularity remains the dominant factor ($POP=0.7$), followed by ND and RI (ND=0.1, RI=0.1).
%This cross-dataset reversal in BPR's behavior underscores the sensitivity of pairwise ranking models to domain-specific data characteristics.

 Comparing across datasets, several patterns emerge. First, Neighborhood Density plays a strong role across nearly all model-dataset combinations, confirming the importance of neighborhood-based signals regardless of domain.
 %By contrast, genre concentration in recipient users' profiles does not predict item exposure, suggesting that genre density is not what the RS optimizes for.
 Second, the Related Items feature shows a clear domain dependency: negligible on ML-1M, it emerges as a significant factor on Last.fm, indicating that item similarity within user profiles carries more weight in the music domain than in the movie domain. This may be due to movie viewers potentially having broader genre preferences, whereas music fans tend to focus more on particular genres. This would lead the RI feature to be more predictive for music. Third, BPR is strongly influenced by item popularity across both datasets, suggesting its global behavior is largely driven by popularity bias. Finally, the acoustic features AE and INS, available only in Last.fm, contribute minimally across all models. It is possible that the information in these features is subsumed by the GD feature.
 %suggesting that content-based attributes play a secondary role compared to structural and collaborative signals, even in a domain where such attributes are readily available.

\begin{figure}[t]
    \centering
    \includegraphics[width=1\columnwidth]{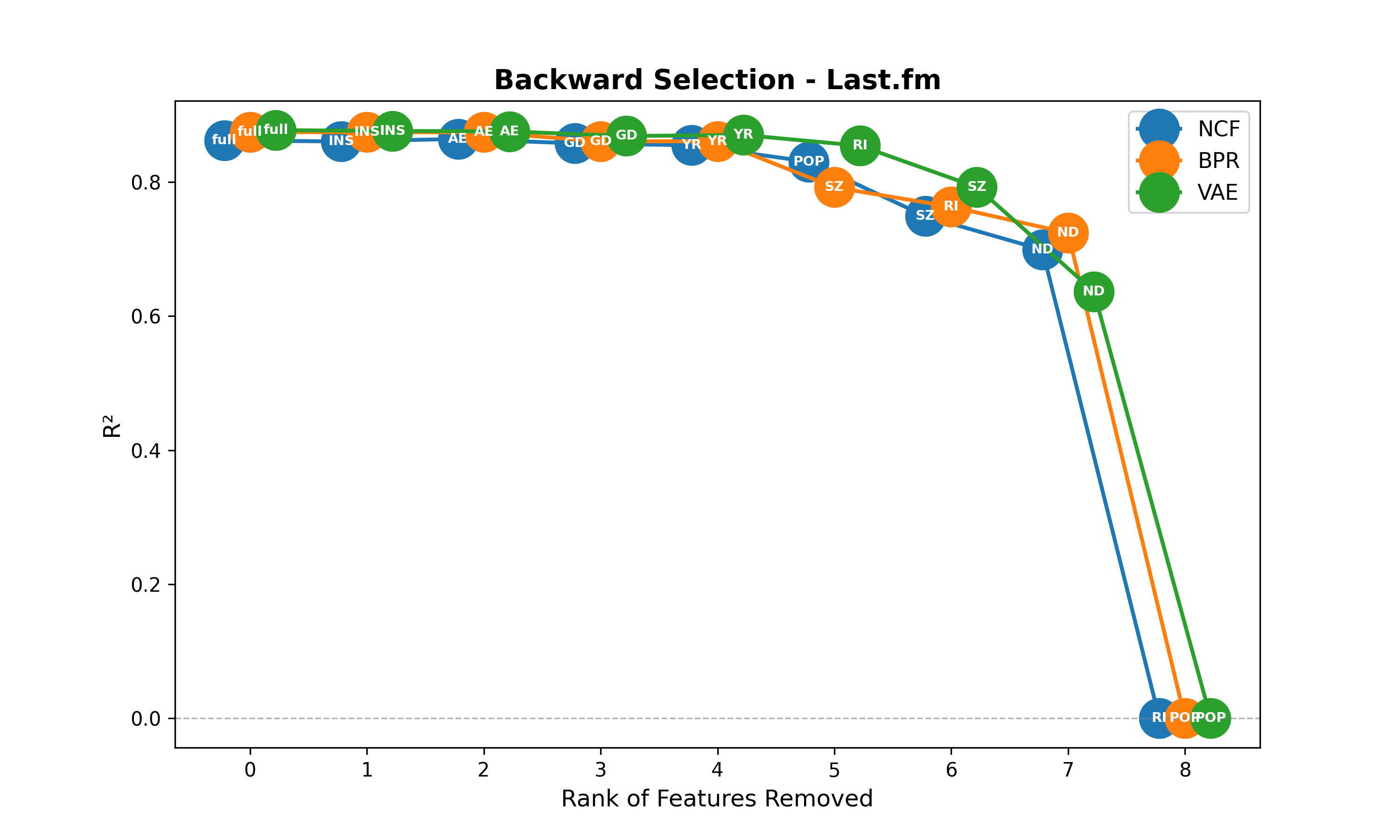}
    \caption{Backward feature elimination results on the Last.fm dataset for NCF, BPR, and VAE models, measured by $R^2$. Each node label indicates the feature removed at that step; rank 0 represents the full feature set.}
    \label{fig:backwardselection}
\end{figure}

Figure~\ref{fig:backwardselection} presents the results of a backward feature elimination study on the Last.fm dataset across three recommender system models, evaluated using $R^2$. At rank 0, all models begin with their full feature set, achieving $R^2\approx0.87$-$0.88$. Features are then removed iteratively in order of least importance, and the resulting $R^2$ is recorded at each step. The curve remains largely flat from ranks 0 to 4, indicating that the early-removed features, INS, AE, GD, and YR, contribute minimally to the model's predictive power across all three models. A gradual decline begins at ranks 5 to 7, where the removal of POP, SZ, and RI introduces moderate performance degradation. The most striking observation occurs at rank 8, where removing the final remaining feature causes a sharp collapse to near zero for all models. This reveals that a single feature alone, RI for NCF and POP for both VAE and BPR, is sufficient to explain $0.6$-$0.7$ of the variance in item exposure.

A key motivation of this work is to provide transparency to item providers. However, it is important to acknowledge that not all the factors identified by our approach are directly actionable by the providers; in this sense, the interpretations we provide are diagnostic rather than prescriptive. Table~\ref{tab:actionability} provides an analysis of the actionability of our feature types from the provider and platform perspectives. However, diagnostic or informative transparency is itself a meaningful and valuable form of explanation~\cite{sonboli2021fairness}. From the provider's perspective, we think it is useful to understand that low exposure may stem from structural dataset properties, such as low neighborhood density or low popularity, rather than item quality alone. This perspective helps providers set realistic expectations, reduces their uncertainty, and replaces speculative folk theories \cite{devito2017algorithms} with evidence-based insight ~\cite{sonboli2021fairness}. From the platform designer's perspective, these findings may be directly actionable: engineers can use this analysis to identify and mitigate structural biases in their recommendation algorithms, for example, by designing re-ranking strategies or post-hoc models to reduce the dominance of neighborhood density or popularity \cite{ aird2025integrating, xu2023p}.

A potential methodological concern is whether the conditioning of features SZ, GD, ND, and RI on $\mathcal{U}_i^*$ introduces circularity. We argue it does not. This conditioning is not incidental but reflects the core explanatory goal of our framework: we seek to characterize the structural properties of recommended items as experienced by the users who actually received them as recommendations. The question we ask is descriptive rather than counterfactual (what properties characterize an item's observed exposure pattern), making conditioning on $\mathcal{U}_i^*$ a deliberate design choice aligned with this post-hoc goal. In addition, we note that Table~\ref{tab:surrogate_results} argues against circularity: GD contributes negligibly across all models and datasets, and RI contributes zero on ML-1M, despite sharing the same conditioning as ND. Meanwhile, POP, entirely independent of $\mathcal{U}_i^*$, dominates for BPR across both datasets.

\section{Conclusion}
We proposed a surrogate modeling framework to explain the black-box recommender systems from the perspective of item providers. Rather than explaining individual user-item pairs, we trained an interpretable proxy model on a set of structural features, including neighborhood density, popularity, profile size, and item similarity, to approximate the global exposure distribution produced by three recommendation models across two domains.
\begin{table}[t]
\centering
\caption{Actionability of features by provider and platform.}
\label{tab:actionability}
\begin{tabular}{lcc}
\toprule
\textbf{Feature} & \textbf{Actionable by provider?} & \textbf{Actionable by platform?} \\
\midrule
YR, AE, INS & \textcolor{green!60!black}{\checkmark} Yes (item properties) & \textcolor{red}{\ding{55}} No \\
\midrule
POP & \textcolor{red}{\ding{55}} No & \textcolor{green!60!black}{\checkmark} Yes (re-ranking) \\
\midrule
ND, RI, SZ, GD & \textcolor{red}{\ding{55}} No & \textcolor{green!60!black}{\checkmark} Yes (algorithm design) \\
\bottomrule
\end{tabular}
\end{table}
Our results demonstrate consistently high surrogate fidelity ($R^2=0.83-0.93$), confirming that structural dataset properties are reliable predictors of recommendation exposure. Feature importance analysis reveals that the dominant drivers vary meaningfully across models and domains: neighborhood density leads exposure for NCF and VAE on ML-1M, popularity dominates  BPR across both datasets, and item similarity plays a notably larger role in the music domain. Acoustic content features contribute minimally.

While not all identified factors are directly actionable by providers, we argue that this distinction does not diminish their value in contributing to explanations. Additional human-centered research is required to understand how the insights available through a global model of this type may be presented as explanations to providers, and to understand how providers would make use of such information.

%%
%% Acknowledgments (uncomment and fill in for the camera-ready version)
%\begin{acknowledgments}
%  ...
%\end{acknowledgments}

%% Required by CEUR-WS since January 2025 — pick/edit the applicable statement.
%% See https://ceur-ws.org/GenAI/Policy.html

%%
%% Bibliography (ceurart handles the style; no \bibliographystyle needed)
\bibliography{PSE}

\end{document}